\documentclass[twocolumn,showpacs,preprintnumbers,amsmath,amssymb]{revtex4}
\usepackage{graphicx}
\usepackage{dcolumn}
\usepackage{bm}

\begin{document}
\title{Fluctuation Cumulant Behavior for the Field-Pulse Induced \\Magnetisation-Reversal Transition in Ising Models}

\author{Arnab Chatterjee}
 \email{arnab@cmp.saha.ernet.in}
\author{Bikas K Chakrabarti}%
 \email{bikas@cmp.saha.ernet.in}
\affiliation{
Saha Institute of Nuclear Physics\\
1/AF Bidhannagar, Kolkata 700 064, India.\\}

\begin{abstract}
The universality class of the dynamic magnetisation-reversal transition,
induced by a competing field pulse, in an Ising model on a square
lattice, below its static ordering temperature, is studied here using
Monte Carlo simulations. Fourth order cumulant of the order parameter
distribution is studied for different system sizes around the phase 
boundary region.
The crossing point of the cumulant (for different system sizes) gives
the transition point and the value of the cumulant at the transition
point indicates the universality class of the transition. 
The cumulant value at the crossing point for low temperature and pulse
width range is observed to be significantly less than that for the static 
transition in the same 
two-dimensional Ising model. The finite size scaling behaviour in this
range also indicates a higher correlation length exponent value. For higher
temperature and pulse width range, the transition seems to fall in a 
mean-field like universality class.
\end{abstract}

\pacs{05.50.+q; 05.70.Fh}
\maketitle

\section {Introduction}
\noindent The response of a pure magnetic system to time-dependent external 
magnetic fields has been of current interest in statistical physics 
\cite{mabkc1,korn1,mish1}. These studies, having close applications in recording
 and switching industry, have also got considerable practical importance.
These spin systems, driven by time-dependent external magnetic fields,
have basically got a competition between two time scales: the time-period
of the driving field and the relaxation time of the driven system.
This gives rise to interesting non-equilibrium phenomena. T\'{o}me
and Oliveira first made a mean-field study \cite{tome} of kinetic
Ising systems under oscillating field. The existence of the dynamic
phase transition for such a system and its nature have been thoroughly
studied using extensive Monte-Carlo simulations. Later, investigations
were extended to the dynamic response of (ferromagnetic) pure Ising
systems under magnetic fields of finite-time duration \cite{pul1}.
All the studies with pulsed field were made below \( T^{0}_{c} \),
the static critical temperature (without any field), where the system
gets ordered. A `positive' pulse is one which is applied along the
direction of prevalent order, while the `negative' one is applied
opposite to that. The results for the positive pulse case did not
involve any new thermodynamic scale \cite{pul1}. In the negative
pulse case, however, interesting features were observed \cite{pul1}:
the negative field pulse competes with the existing order, and the system
makes a transition from one ordered state characterised by an equilibrium
magnetisation \( +m_{0} \) (say) to the other equivalent ordered
state with equilibrium magnetisation \( -m_{0}, \) depending on the
temperature \( T \), field strength \( h_{p} \) and its duration
\( \Delta t \). This transion is well studied in the limit \( \Delta t\rightarrow \infty  \)
for any non-zero value of \( h_{p} \) at any \( T<T_{c}^{0} \).
This transition, for the general cases of finite \( \Delta t \),
is called here `magnetisation-reversal' transition. Some aspects of this
transition has been recently studied extensively \cite{mish1,pul2}.

\section {Model and the Transition}
The model studied here is the Ising model with nearest-neighbour interaction
under a time-dependent external magnetic field. This is described
by the Hamiltonian:
\begin{equation}
\label{ham}
H=-\frac{1}{2}\sum _{\{ij\}}J_{ij}S_{i}S_{j}-h(t)\sum _{i}S_{i},
\end{equation}

\noindent where \( J_{ij} \) is the cooperative interaction between
the spins at site \( i \) and \( j \) respectively, and each nearest-neighbour
pair is denoted by \{...\}. We consider a square lattice. The static
critical temperature is \( T_{c}^{0}=2/\ln (1+\sqrt{2})\simeq 2.269... \) (in units
of \( J/K_{B} \)). At \( T<T_{c}^{0} \), an external field pulse
is applied, after the system is brought to equilibrium characterised
by an equilibrium magnetisation \( m_{0}(T) \). The spatially uniform
field has a time-dependence as follows:\begin{equation}
\label{hpdef}
h(t)=\left\{ \begin{array}{cc}
-h_{p} & t_{0}\leq t\leq t_{0}+\Delta t\\
0 & \rm otherwise.
\end{array}\right. 
\end{equation}

Typical time-dependent (response) magnetisation \( m(t) \) (= \( <S_{i}>, \)
where $<...>$ denotes the thermodynamic `ensemble' average) of the system
under different magnetic field \( h(t) \) are indicated in the Fig. 1.
The time \( t_{0} \) at which the pulse is applied is chosen such
that the system reaches its equilibrium at \( T \) ($<$ \( T_{c}^{0} \)
). As soon as the field is applied, the magnetisation \( m(t) \)
starts decreasing, continues until time \( t+\Delta t \) when the
field is withdrawn. The system relaxes eventually to one of the two
equlibrium states (with magnetisation \( -m_{0} \) or \( +m_{0} \)).
At a particular temperature \( T \), for appropriate combinations
of \( h_{p} \) and \( \Delta t, \) a magnetisation-reversal transition
occurs, when the magnetisation of the system switches from one state
of equilibrium magnetisation \( m_{0} \) to the other with magnetisation
\( -m_{0} \). This reversal phenomena at \( T<T_{c}^{0} \) is simple
and well studied for \( \Delta t\rightarrow \infty  \) for any non-zero
\( h_{p} \). We study here the dynamics for finite \( \Delta t \)
values. It appears that generally \( h_{p} \rightarrow \infty \) as
\( \Delta t\rightarrow 0 \) and \( h_{p} \rightarrow 0 \) as 
 \( \Delta t\rightarrow \infty  \) for any such dynamic magnetisation-reversal
transition phase boundary at any temperature \( T \) ($<$ \( T_{c}^{0} \)).
In fact, a simple application of the domain nucleation theory gives 
\(h_p \rm{ln} \Delta t\) = constant along the phase boundary, where the constant
changes by a factor \(1/(d+1)\), where \(d\) denotes the lattice dimension,
 as the boundary changes from single to multi-domain region \cite{pul1}. 
\vskip 0.5in

{\centering \resizebox*{8cm}{6.5cm}{\rotatebox{0}{\includegraphics{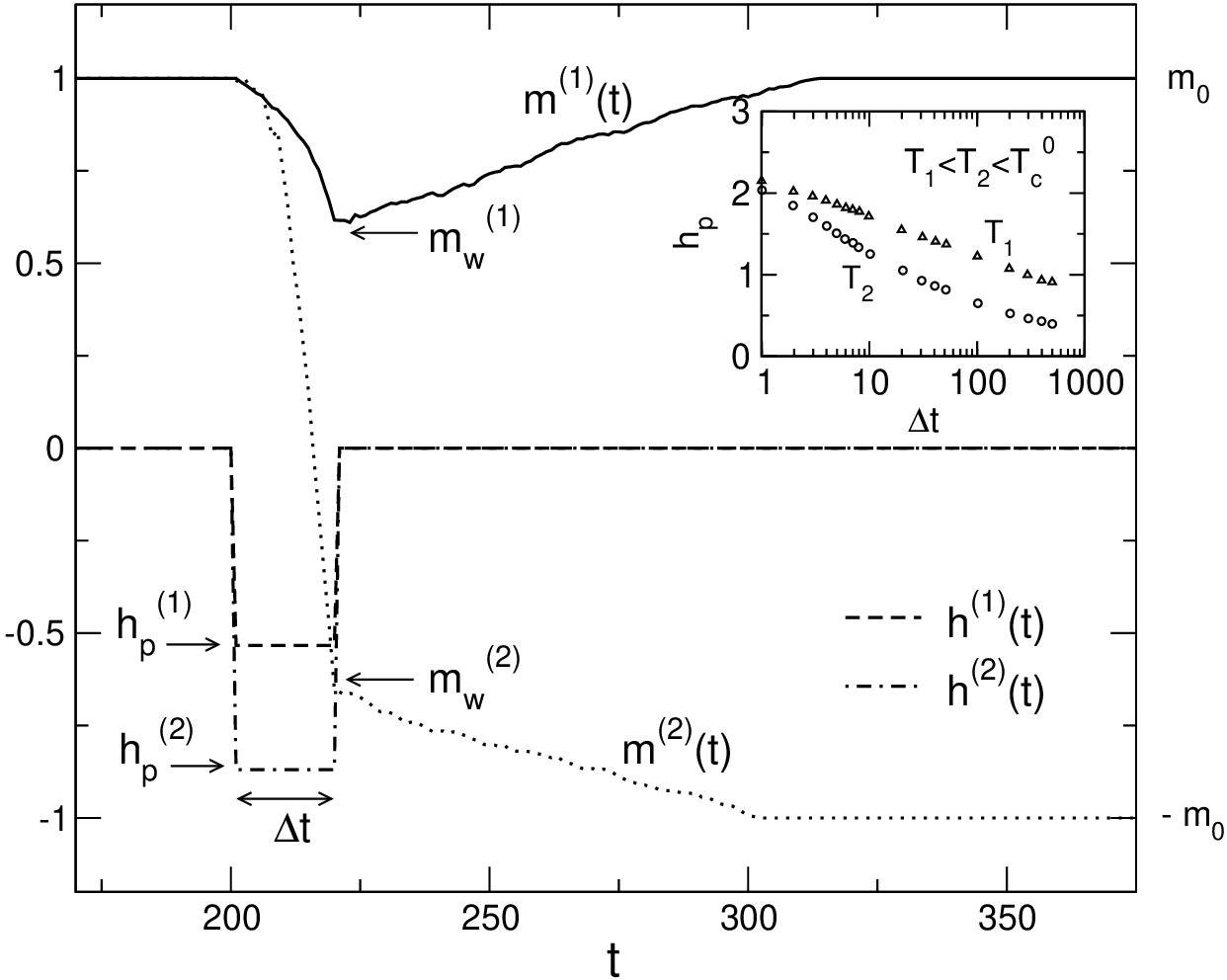}}} \par}
\vspace{0.3cm}
\vskip 0.1in
\noindent {\footnotesize FIG. 1. {Typical time variation of the response 
magnetisation \( m(t) \) for two different field pulses \( h(t) \) with same 
\( \Delta t \) for an Ising system at a fixed temperature \( T \).  The 
magnetisation-reversal here occurs due to increased pulse strength, keeping 
their width \( \Delta t \) same. The transition can also be brought about by 
increasing \( \Delta t \), keeping \( h_{p} \) fixed. The inset indicates the 
typical phase boundaries (where the field withdrawal-time magnetisation 
\( m_{w}=0 \)) for two different temperatures (sequential updating; note that 
for random updating the phase boundaries shift upwards).}}{\footnotesize \par}
\vskip 0.1in

A mean field study of the problem gives a qualitative understanding
of the diverging time and length scales developed near the transition
boundary (in the \( h_{p} \) - \( \Delta t \) plane at a fixed \( T<T^{0}_{c} \)).
Mean-field approximation for the dynamics gives the equation of motion
for the average magnetisation \( m_{i} \) as \begin{equation}
\label{motion}
\frac{dm_{i}}{dt}=-m_{i}+\tanh \left( \frac{\sum _{j}J_{ij}m_{j}+h(t)}{T}\right).
\end{equation}

\noindent This equation, linearised near the magnetisation-reversal
transition point, gives, for uniform magnetisation, \begin{equation}
\label{solmf}
m(t)=\frac{h_{p}}{\Delta T}-\left( \frac{h_{p}}{\Delta T}-m_{0}\right) \left[ \exp \left\{ \frac{\Delta T}{T}(t-t_{0})\right\} \right]
\end{equation}

\noindent as a solution of eqn. (\ref{motion}), for \( t_{0}\leq t\leq t_{0}+\Delta t \). Here \( \Delta T=T^{mf}_{c}-T \),
where \( T^{mf}_{c}\equiv J(q=0) \) is the static critical temperature in
the mean-field approximation and \( J(q) \) is the Fourier transform
of the interaction \( J_{ij} \). Due to application of the field \( h_{p}, \)
\( m(t) \) decreases in magnitude from \( m(t_{0})\equiv m_{0} \)
to \( m(t_{0}+\Delta t)\equiv m_{w} \) at the time of withdrawal
of the pulse. Due to absence of fluctuation here, magnetisation relaxes
back to its original value \( m_{0} \) if \( m_{w} \) is positive,
or to a value \( -m_{0} \) if \( m_{w} \) is negative. 
In the \( t>t_{0}+\Delta t \) regime, where \( h(t)=0, \)
the magnetisation (starting from \( m_{w} \) at \( t=t+\Delta t \))
relaxes back to its final equilibrium value \( \pm m_{0} \), with
a relaxation time \cite{mish1,pul1}
\begin{equation}
\label{tau}
\tau \sim \frac{1}{(T_{c}^{mf}-T)}\ln \left( \left| \frac{m_{0}}{m_{w}}\right| \right) .
\end{equation}

\noindent It diverges at the magnetisation-reversal transition boundary,
where \( m_{w} \) vanishes. The prefactor gives the divergence of
\( \tau  \) at the static mean field transition temperature, and
is responsible for critical slowing down phenomena at the static transition
point (\( h=0 \)). The other factor gives the diverging time scale,
at any temperature below the static transition temperature, where
magnetisation reversal occurs or \( m_{w} \) vanishes due to appropriate
combination of \( h_{p} \) and \( \Delta t. \) The solution of the
susceptibility \( \chi (q) \) gives \cite{mish1} \begin{equation}
\label{chi}
\chi (q)\sim \exp \left( -q^{2}\xi ^{2}\right) ,
\end{equation}

\noindent where the correlation length is given by

\begin{equation}
\label{xi}
\xi \sim \left[ \frac{1}{(T_{c}^{mf}-T)}\ln \left( \left| \frac{m_{0}}{m_{w}}\right| \right) \right] ^{\frac{1}{2}}.
\end{equation}

\noindent Here, too, the prefactor in \( \chi  \) gives the usual
divergence at \( T^{mf}_{c}, \) while the other factor gives the
divergence at the magnetisation-reversal transition point. Incorporating
fluctuations, extensive Monte-Carlo simulation studies have also convincingly
demonstrated \cite{pul2} that the fluctuation in the order parameter 
\( |m_{w}| \) and
in the internal energy of the system grows with the system size and
diverges at the magnetisation-reversal transition boundary, where
\( m_{w} \) vanishes.

\section {Monte-Carlo study and the results}
Here the Monte-Carlo study has been carried out in two-dimensions
(square lattice) with periodic boundary conditions. Spins are updated
following Glauber dynamics. The updating rule employed here are both
sequential as well as random. In sequential updating rule one Monte-Carlo
step consist of a complete scan of the lattice in a sequential manner;
while in random updating a Monte-Carlo step is defined by \( N \)
(= \( L^{2} \) ) random updates on the lattice, where \( N \) is
the total number of spins in a lattice of linear size \( L \). Studies
have been carried out at temperatures below the static critical temperature
(\( T_{c}^{0}\simeq 2.27 \)). The system is allowed to evolve from an initial
state of perfect order to its equilibrium state at temperature \( T \). 
The time \( t_{0} \) is chosen to be
much larger than the static relaxation time at that \( T \), so that
the system reaches an equilibrium state with magnetization \( +m_{0}(T) \)
before the external magnetic field is applied at time \( t=t_{0} \).
The field pulse of strength \( -h_{p} \) is applied for duration
\( \Delta t \) (measured in Monte Carlo steps or MCS). The magnetisation
starts decreasing from its equilibrium value \( m_{0} \). The average
value of the magnetisation \( m_{w} \) at the time of withdrawal
of pulse is noted. The phase boundary of this dynamic transition is
defined by appropriate combination of \( h_{p} \) and \( \Delta t \)
that produces the magnetisation reversal by making \( m(t_{0}+\Delta t)\equiv m_{w}=0 \)
from a value \( m(t_{0})=m_{0}, \) i.e, \( m_{w} \) changes sign
across the phase boundary. The phase boundary changes with \( T. \)
The bahavior of different thermodynamic quantities are studied across
the phase boundary. These quantities are averaged over \( 1000-20000 \) 
different initial configurations of the system. The fluctuations over 
the average value are also noted. 

Here we study the behavior of the reduced fourth order cumulant \( U \)
\cite{binheer} near the magnetisation reversal transition. This is
defined as \begin{equation}
\label{bind}
U=1-\frac{\left\langle m_{w}^{4}\right\rangle }{3\left\langle m_{w}^{2}\right\rangle ^{2}},
\end{equation}

\noindent where \( \left\langle m_{w}^{4}\right\rangle  \) is the
ensemble average of \( m_{w}^{4} \). \( \left\langle m_{w}^{2}\right\rangle  \)
is similarly defined. The cumulant \( U \) here behaves 
somewhat differently, compared to that in static and other transitions: Deep
inside the ordered phase \( m_{w}\simeq 1 \) and \( U\rightarrow 2/3. \)
For other (say, static) transitions the order parameter (\( m_{w} \))
goes to zero with a Gaussian fluctuation above the transition point,
 giving \( U\rightarrow 0 \) there. Here,
however, due to the presence of the pulsed field, \( \left| m_{w}\right|  \)
is non-zero on both sides of the magnetisation-reversal transition.
Hence \( U \) drops to zero at a point near the transition and grows again
after it.

The universality class of the dynamic transition in Ising model under
oscillating field has been studied extensively by investigating \cite{korn1}
the critical point and the cumulant value \( U^{*} \) at the critical
point, where the cumulant curves cross for different system sizes
(\( L \)). In that case, of course, the variation of \( U \) (at
any fixed \( L \)) is similar to that in the static Ising transitions
(\( U=2/3 \) well inside the ordered phase and \( U\rightarrow 0 \)
well within the disordered phase). In fact, \( U^{*} \) value in this 
oscillatory field case was found to be the same as that in the static case, 
indicating the same universality class \cite{korn1}. We observe different
bahaviour in the field pulse induced magnetisation-reversal transition case.

We observe two kinds of distinct behavior of the cumulant \( U. \)
Typically, for low temperature and low pulse-duration region (see
the inset in Fig. 1) of the magnetisation-reversal phase boundary,
the cumulant crossing for different system sizes (\( L \)) occur
at \( U^{*}\simeq 0.42 \) to \( 0.46 \) (see Fig. 2). As mentioned already, we
have checked these results for both sequential and random updating. 
Specifically, for \( T=0.5 \) and \( \Delta t=5, \) (see Fig. 2c) we find the 
transition point value of \( h_{p}\simeq 2.6, \) to be smaller than the value 
(\( \simeq 1.9 \)) for sequential updating. However, the value of \( U^{*} \) 
at this transition point is again very close to about \( 0.44. \) This indicates
that updating rule does not affect the universality class (\( U^{*} \)
value), as long as the proper region of the phase boundary is considered.
For relatively higher temperature and pulse-duration region of the
phase boundary, the crossing of \( U \) for different \( L \) values
occur for \( U^{*}\simeq 0_{+} \). This is true for both sequential (Fig. 
3a, b) and random (Fig. 3c) updating. It may be noted that the phase
boundary changes with the updating rule, as the system relaxation
time (which matches with the pulse width at the phase boundary) is
different for sequential and random updating \cite{binheer}.

\vskip 0.4in
{\centering \resizebox{7.5cm}{6.0cm}{\rotatebox{0}{\includegraphics{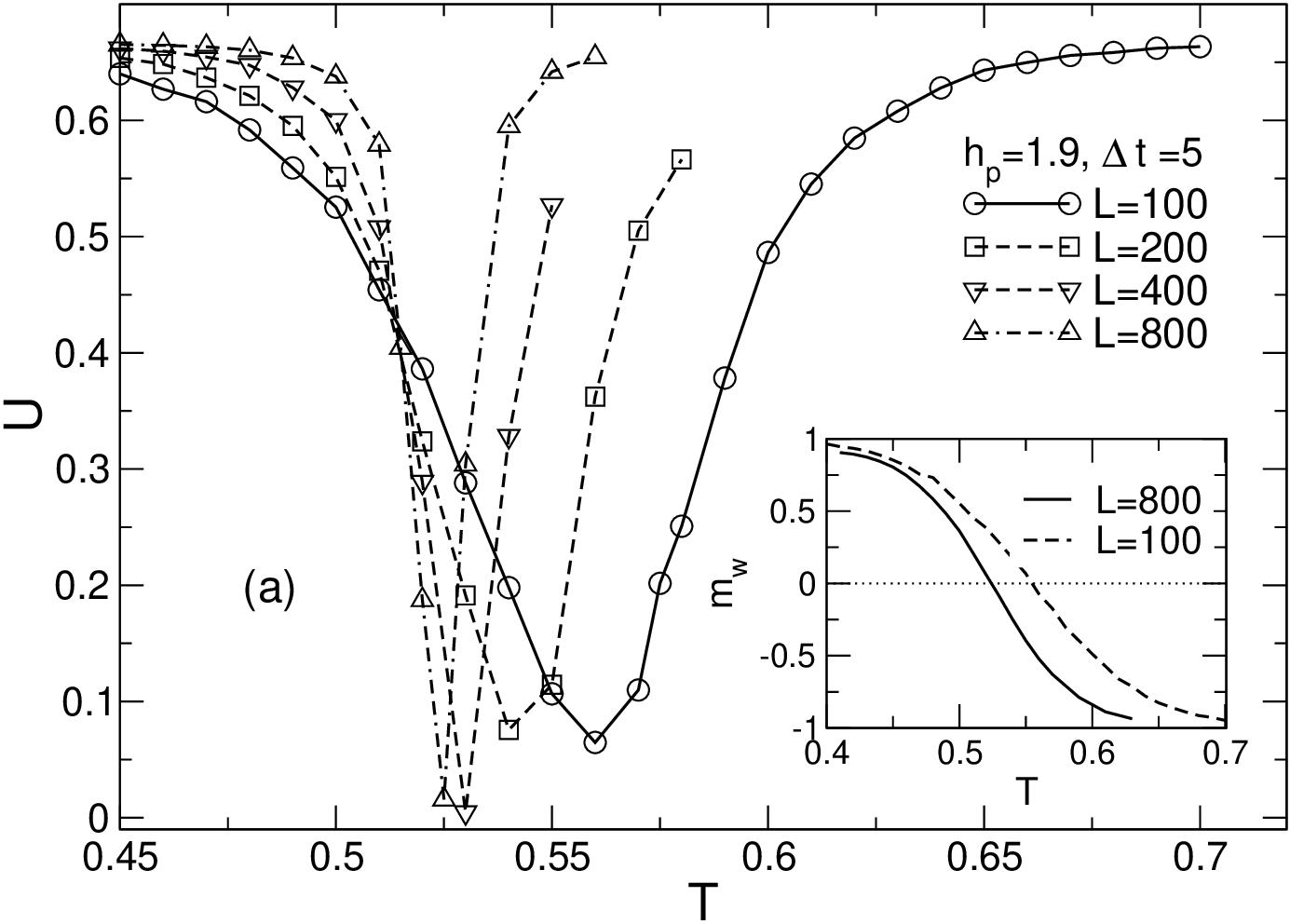}}}}
\vskip 0.3in
{\centering \resizebox{7.5cm}{6.0cm}{\rotatebox{0}{\includegraphics{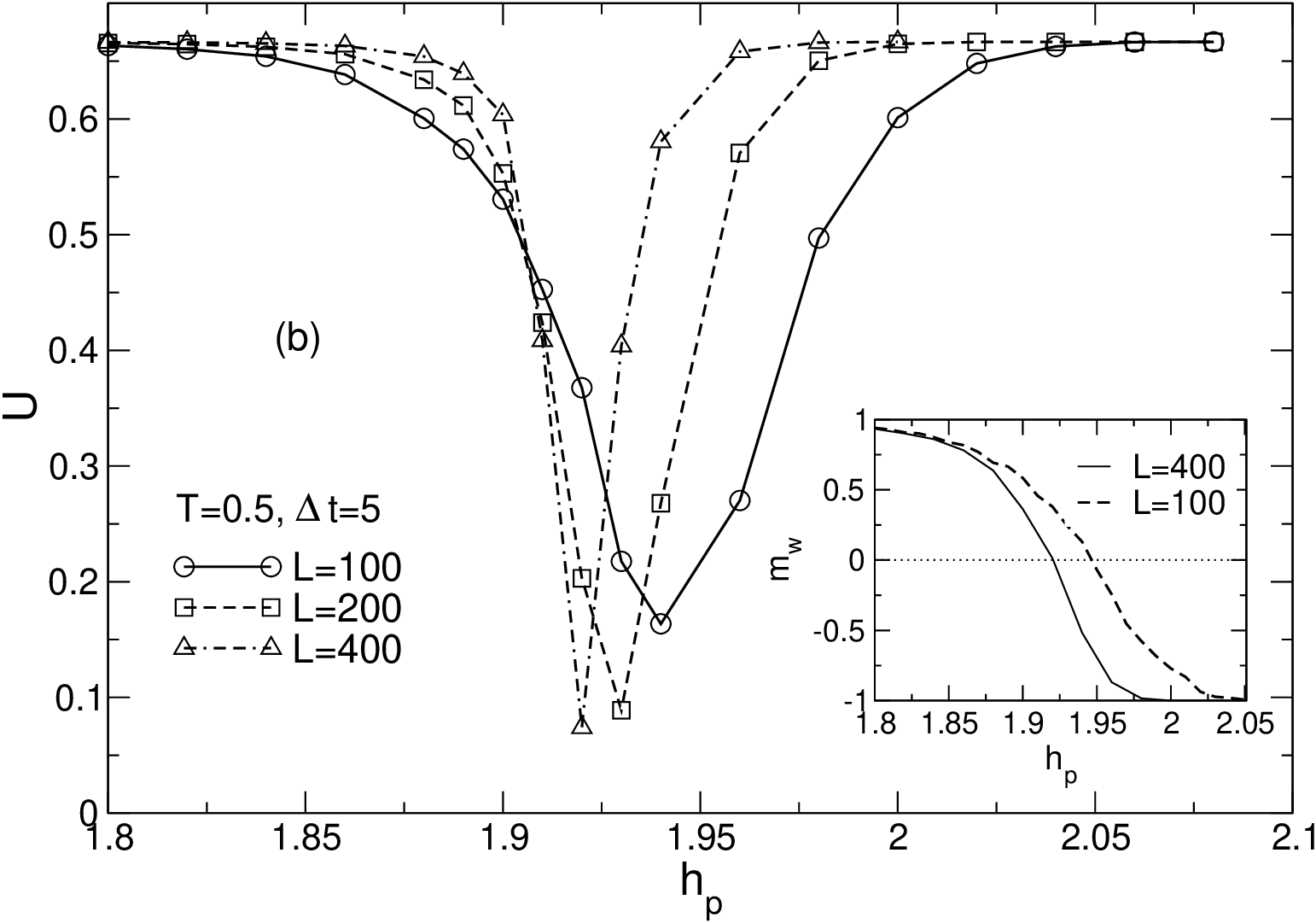}}}} 
{\centering \resizebox{7.5cm}{6.0cm}{\rotatebox{0}{\includegraphics{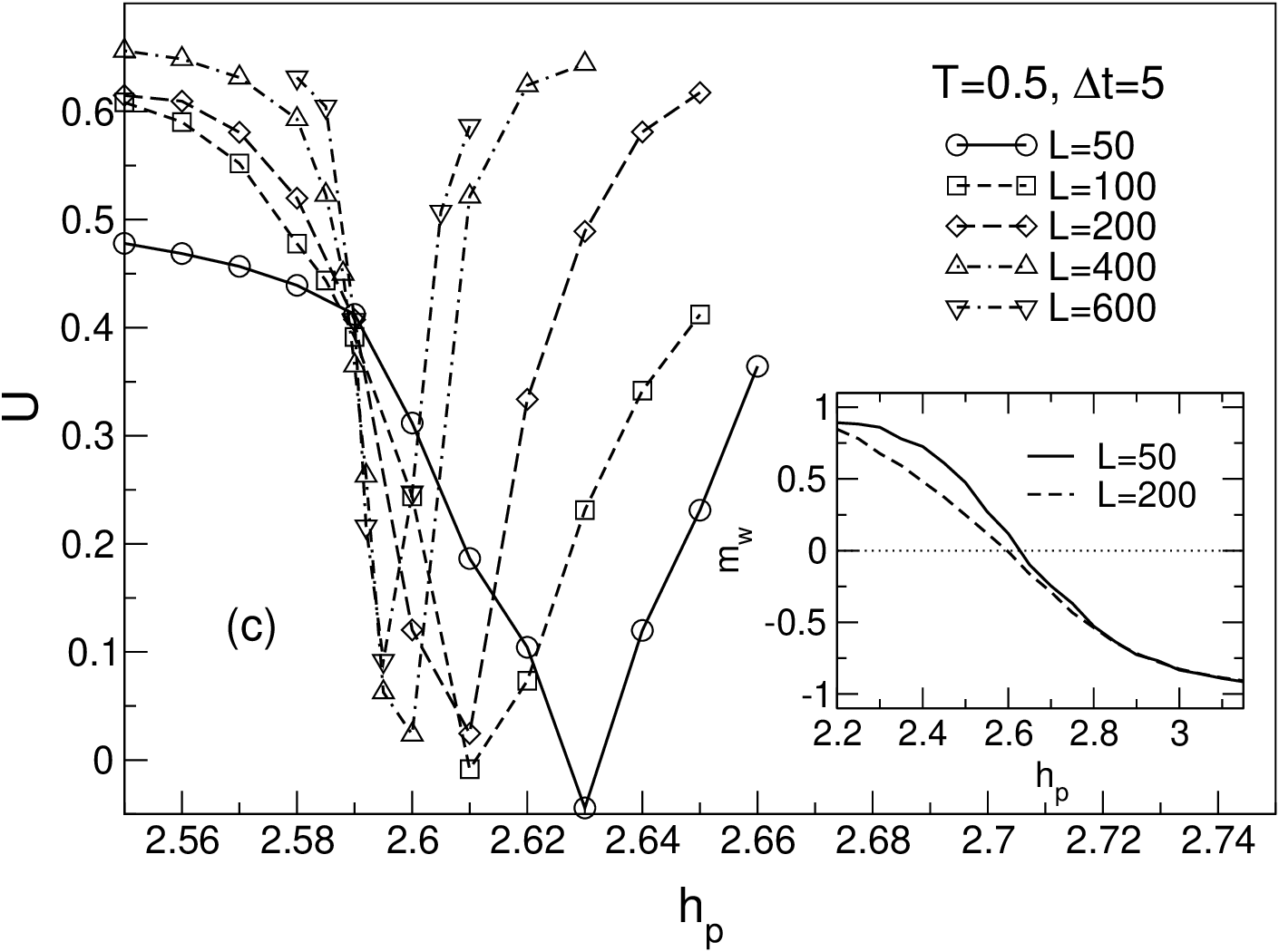}}}}
\vskip 0.4in
{\centering \resizebox{7.5cm}{6.0cm}{\rotatebox{0}{\includegraphics{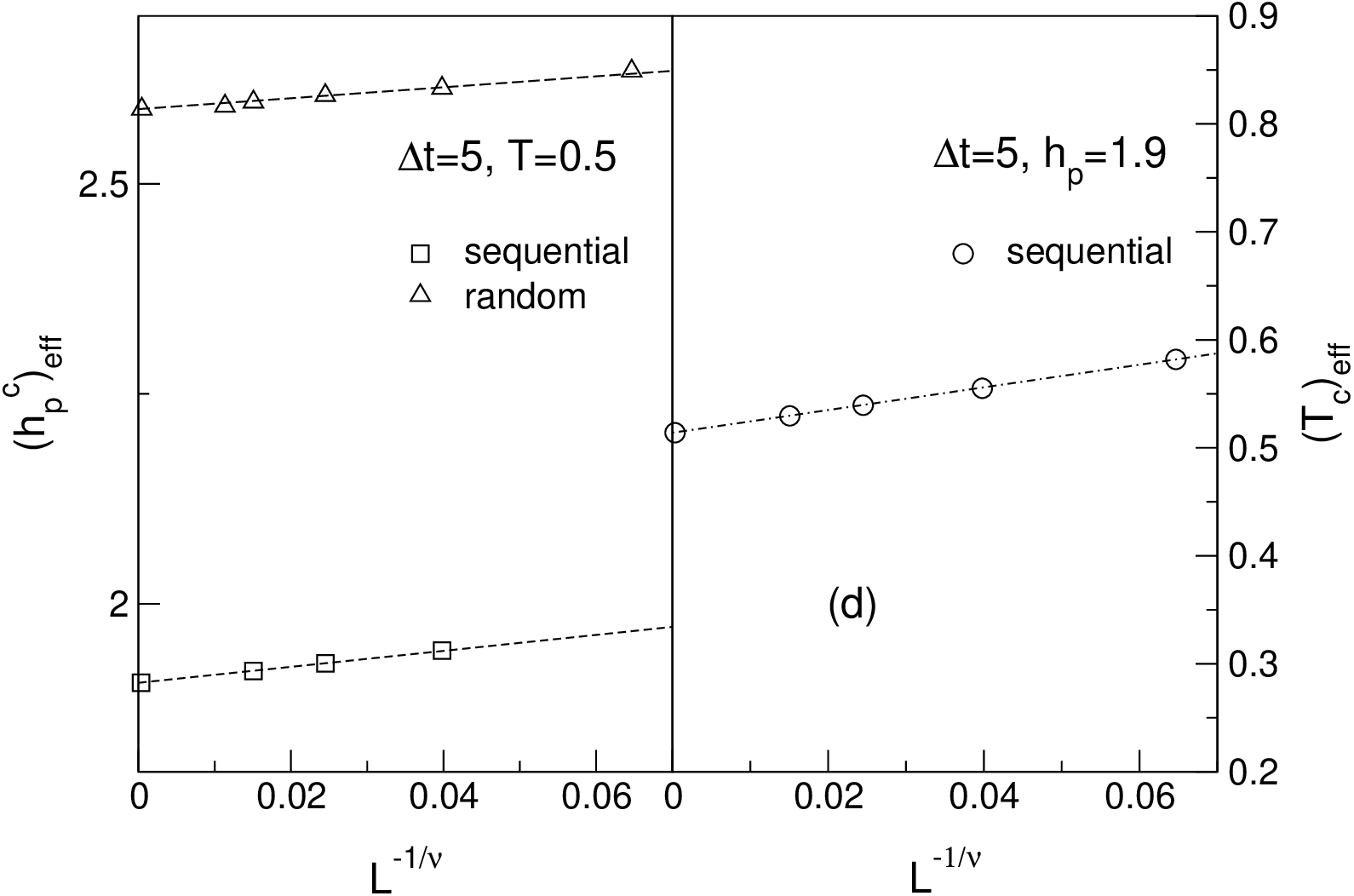}}}} 
\vskip 0.1in

\noindent {\footnotesize FIG. 2. 
 Behavior of \( U \) near the transition,
driven by (a) \( T \) at a fixed value of \( h_p \) (=1.9) and \( \Delta t \)
(=\( 5 \)) with sequential updating, (b) \( h_p \) at a fixed value of \( T \)
(=0.5) and \( \Delta t \) (=\( 5 \)) with sequential updating, and 
(c) \( h_{p} \) at a fixed value of \( T \) (=0.5) and \( \Delta t \)
(=\( 5 \)) with random updating,
 for different \( L \), averaged over 1000 to 20000 initial
configurations.
The fluctuations are smaller than the symbol size. The insets show the typical 
behavior of the magnetisation \( m_{w} \) at the time of withdrawal of the
field pulse by varying (a) \( T \) at a fixed \( h_p \) and \( \Delta t \),
for \( L=100\) and \(800 \), (b) \( h_p \) at a fixed
\( T \) and \( \Delta t \), for \( L=100\) and \( 400 \), (c) \( h_{p} \) at a fixed \( T \) and \( \Delta t \),
for \( L=50\) and \( 200 \); \( m_{w}=0 \) at the effective transition point.
(d) Finite size scaling study in this parameter range: the effective \( T_c \)
or \( h_{p}^c \) values (see the insets), where \( m_{w}=0 \), are plotted 
against \(L^{-1/\nu}\) with \(\nu^{-1}=0.7\). The values of the cumulant 
crossing points in (a), (b), (c) are taken to correspond the respective 
transition points for \(L \rightarrow \infty\).
}{\footnotesize \par}
\vskip 0.1cm

It might be noted that in the low temperature and \( \Delta t \)
regions, there seems to be significant finite size scaling of the
transition (\( m_{w}=0) \) point (see the insets of Fig. 2a, b, c).  In fact, 
in Fig. 2d, the finite-size scaling analysis of those data is presented.
For the other cases, there seems to be no significant finite size effect 
on the transition point (cf. insets of Fig. 3a, b, c), indicative of 
a mean-field nature of the transition in this range.
It may be noted that to compare the finite size effects, we normalise the
parameters \(T\) or \(h_p\) by their ranges required for full magnetisation
reversal. In fact, this weak finite size effect for high \(T\) and
\(\Delta t\) regions did not lead to any reasonable value for the fitting 
exponent in the scaling analysis.

\vskip 0.4in

{\centering \resizebox{7.5cm}{6.0cm}{\rotatebox{0}{\includegraphics{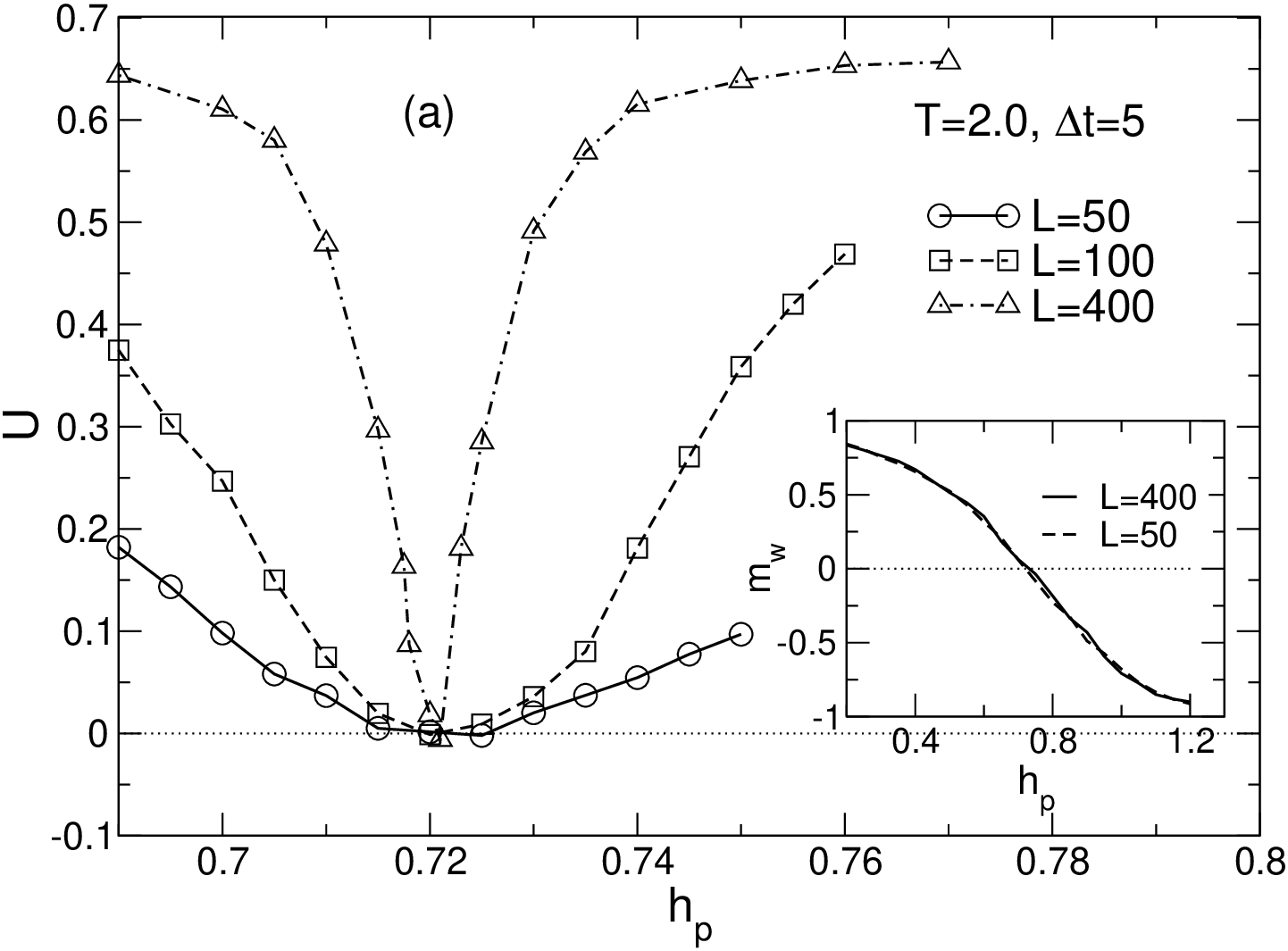}}}}
\vskip 0.2in
{\centering \resizebox{7.5cm}{6.0cm}{\rotatebox{0}{\includegraphics{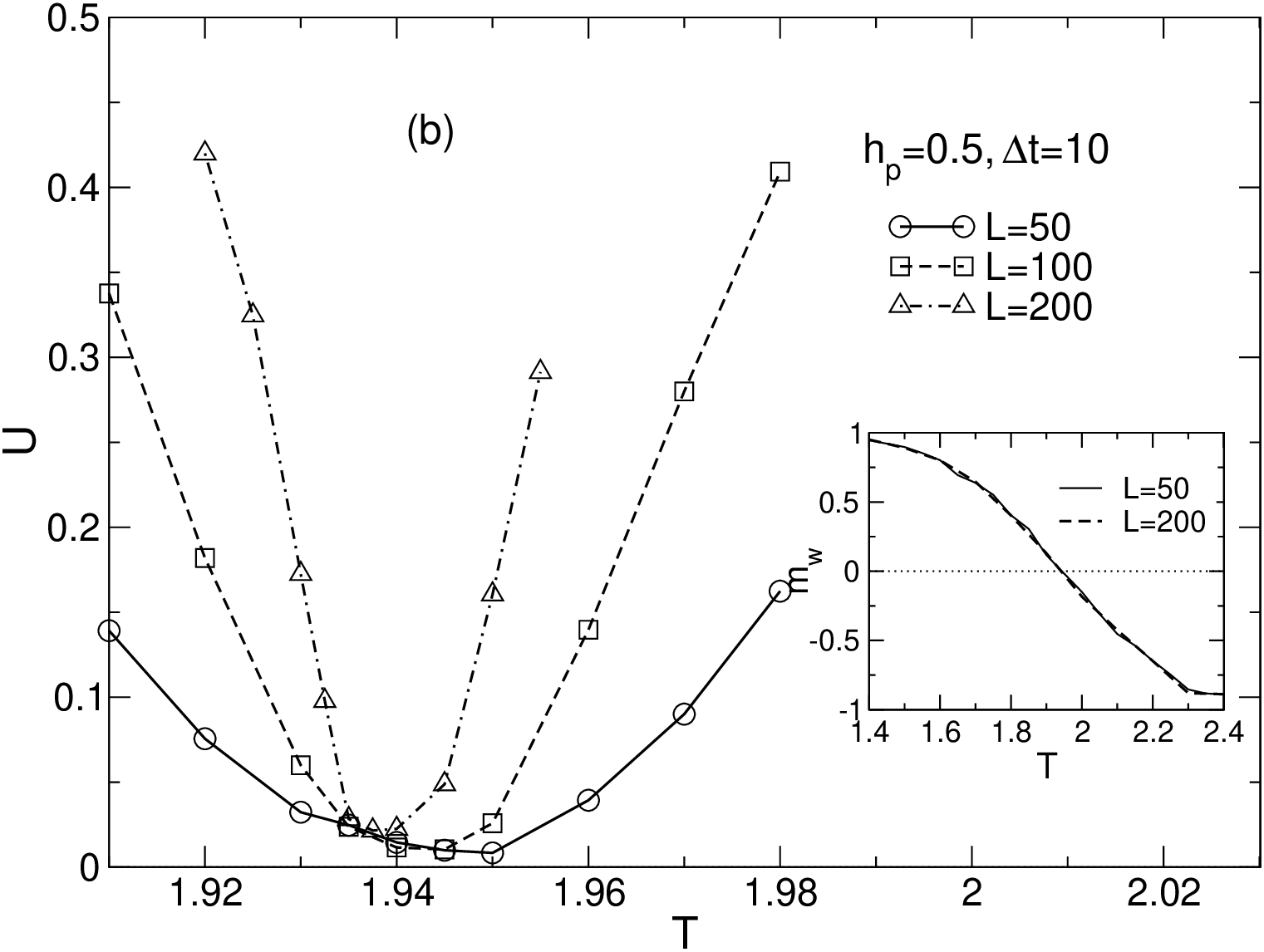}}}} 
\vskip 0.4in
{\centering \resizebox*{8cm}{6cm}{\rotatebox{0}{\includegraphics{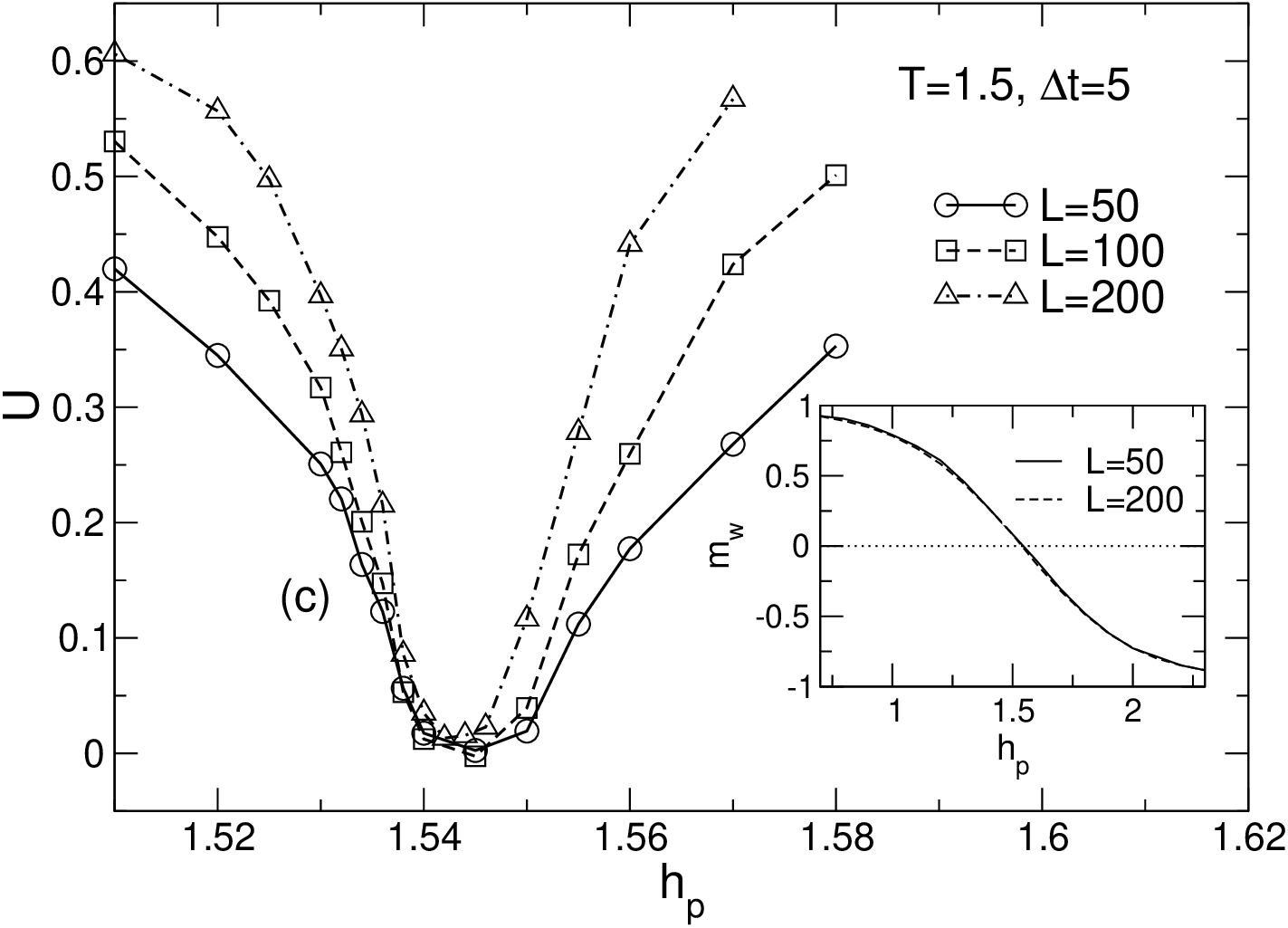}}} \par}
\vskip 0.2in

\noindent {\footnotesize FIG. 3. 
 Behavior of \( U \) near the transition,
driven by (a) \( h_{p} \) at a fixed value of \( T \) (=2.0) and \( \Delta t \)
(=\( 5 \)) with sequential updating, (b) \( T \) at a fixed value of \( h_p \)
(=0.5) and \( \Delta t \) (=\( 10 \)) with sequential updating, and 
(c) \( h_{p} \) at a fixed value of \( T \) (=1.5) and \( \Delta t \)
(=\( 5 \)) with random updating,
 for different \( L \), averaged over 1000 to 6000 initial
configurations.
The fluctuations are smaller than the symbol size. The insets show the typical 
behavior of the magnetisation \( m_{w} \) at the time of withdrawal of the
field pulse by varying (a) \( h_{p} \) at a fixed \( T \) and \( \Delta t \),
for \( L=50\) and \( 400 \), (b) \( T \) at a fixed
\( h_{p} \) and \( \Delta t \), for \( L=50\) and \(200 \), (c) \( h_{p} \) 
at a fixed \( T \) and \( \Delta t \),
for \( L=50\) and \(200\); \( m_{w}=0 \) at the transition point.
}{\footnotesize \par}
\vskip 0.1cm


For the static transition of the pure two-dimensional Ising system, 
\( U^{*}\simeq 0.6107 \) \cite{binheer,blote,bincu}. 
 For low temperature (and low \( \Delta t \)) regions of the 
magnetisation-reversal phase boundary, the observed values of \( U^{*} \) (in 
the range 0.42 - 0.46) are considerably lower than the above mentioned value 
for the static transition. There is not enough indication of finite-size 
effect in the \( U^{*} \) value either (cf. \cite{korn1}). This suggests a new
universality class in this range. Also, the finite-size 
scaling study for the effective transition points here (see Fig. 2d) 
gives a correlation length exponent value (\(\nu \simeq  1.4\)) larger
than that of the static transition.  For comparatively 
higher temperatures (and high \( \Delta t \)), 
the \( U^{*}\simeq 0_{+} \) at the crossing point. Such small value
of the cumulant at the crossing point can hardly be imagined to be a
finite-size effect; it seems unlikely that one would get here also
the same universality class and \( U^{*} \) value will eventually shoot
up to \( U^{*}\simeq 0.44 \) (for larger system sizes), as for the other
range of the transition. On the other hand, such low value of \( U^{*} \)
might indicate a very weak singularity, as indicated by the mean field
calculations \cite{mish1} mentioned in the introduction. In fact,
even for the static transition, as the dimensionality increases, and
the singularity becomes weaker (converging to mean field exponents)
with increasing lattice dimension, the cumulant crossing point \( U^{*} \)
decreases (\( U^{*}\simeq 0.61 \) in \( d=2 \) to \( U^{*}\simeq 0.44 \)
in \( d=4 \)) \cite{bincu}. We believe the mean field transition
behavior here, as mentioned earlier, is even weaker in this dynamic
case as reflected by the value \( U^{*}\simeq 0_{+} \), corresponding
to a logarithmic singularity (as in eqns. (\ref{tau}) and (\ref{xi})).

\section {Summary and conclusions}

The universality class of the dynamic magnetisation-reversal transition,
induced by a competing pulse, in an Ising model on a square lattice, below
its static ordering temperature, 
is studied here using Monte Carlo simulations. Both sequential and
random updating have been used. The phase boundary at any \( T \)
(\(< T_{c}^{0} \)) is obtained first in the \( h_{p}-\Delta t \)
plane. They of course depend on the updating rule. 
The phase boundaries obtained compare well with the nucleation theory
estimate \(h_p \rm{ln} \Delta t\) = constant along the boundary \cite{pul1}.
The mean-field theory applications \cite{mish1,pul1} indicated time 
 and length (eqns. \ref{tau} and \ref{xi} respectively) scale divergences 
at these phase boundaries. Extensive 
Monte-Carlo studies for the fluctuations in the order parameter \(|m_w|\)
and internal energies etc. showed prominent divergences along the phase
boundaries \cite{pul2}. Fourth order cumulant
(\( U \)) of the order parameter distribution is studied here for different
system sizes (upto \( L=800 \)) around the phase boundary region.
The crossing point of the cumulant (for different system sizes) gives
the transition point and the value \( U^{*} \) of the cumulant at
the transition point indicates the universality class of the transition.
In the low temperature and low pulse width range, the \( U^{*} \)
value is found to be around 0.44 (see Figs. 2a, b, c). The prominent discripancy
with the \( U^{*} \) value (\(\simeq 0.61\)) for the static transition in the
same model in two dimensions indicates a new universality class for this 
dynamic transition. Indeed, the finite-size scaling analysis (Fig. 2d) 
suggests a different (larger) value of the correlation length exponent also.
For comparatively higher
temperatures and higher pulse widths, the \( U^{*} \) values are very close
to zero (see Fig. 3a, b, c), and the transitions here seem to fall in a 
mean-field-like
weak-singularity universality class similar to that obtained earlier
\cite{mish1}, and indicated by eqns. (\ref{tau}) and (\ref{xi}).
Here, the finite size effects in the order parameter and the transition
point are also observed to be comparatively weaker (see insets of Fig. 3).

\section {Acknowledgements}

The authors are grateful to Bernard Nienhuis for some of his insightful
questions regarding the transition, and to
Arkajyoti Misra for useful comments and discussions.


\newpage
\end{document}